\documentclass[3p,times]{elsarticle}

\usepackage{ecrc}

\volume{00}

\firstpage{1}

\journalname{Results in Engineering}

\runauth{}

\jid{procs}

\jnltitlelogo{Results in Engineering}

\CopyrightLine{2011}{Published by Elsevier Ltd.}

\usepackage{amssymb}

\usepackage[figuresright]{rotating}

\usepackage{subfigure}
\usepackage{svg}

\begin{document}

\begin{frontmatter}



\dochead{}

\title{Preliminary Study of the Effects of Leading-Edge Serration on a Two-Section Planar Wing in ground-effect at Low Reynolds Number}


\author[label1]{Lafond-Saunier, A.}
\author[label2]{Basile, S.}
\author[label1]{Pizarro, P.}
\author[label3]{Yamamoto, K.}
\author[label3]{Nagib, H.M.}
\author[label1,label4]{Vinuesa, R.}
\author[label1]{Mariani, R.}
\address[label1]{KTH Royal Institute of Technology, Dpt. of Engineering Mechanics, Stockholm, 100 24, Sweden}
\address[label2]{Politecnico di Torino, Dpt. of Mechanical and Aerospace Engineering, Torino, 101 29, Italy}
\address[label3]{Illinois Institute of Technology, Armour College of Engineering, Chicago, IL, 60616, USA}
\address[label4]{University of Michigan, Dpt. of Aerospace Engineering, Ann Arbor, MI; 48109, USA}

\begin{abstract}
A preliminary study has been conducted on the effects of serration on the leading-edge of a two-element trapezoidal wing placed both out-of- and in-ground effect. Aerodynamic performance and flow behaviour were evaluated numerically and validated experimentally. Results indicate an increase in maximum lift coefficient and stall angle obtained implementing a serrated leading-edge geometry due to the flow being re-energized by the formation of a series of counter-rotating pairs of vortices.

Results from the analysis of the wing in ground effect appear less well defined. Both leading-edge geometries -- straight and serrated -- show an increase in efficiency due to the proximity to the ground. The wing with the straight leading-edge geometry shows constant improvement up to stall, whilst numerical results show a significant decrease in lift performance at high angles of attack. This may be caused by the lower-fidelity numerical model implemented at higher angles of attack, thus yielding less accurate results.
\end{abstract}


\end{frontmatter}


\section{Introduction}\label{Sec1}
Flying in ground-effect has the advantage of reducing the strength of wing-tip vortices and reducing downwash \cite{Toosi2024}, resulting in an overall increase in aerodynamic efficiency of a wing \cite{Borst1979,Yun2010}. Two design philosophies, sharing the low-aspect ratio wing concept, have been implemented to exploit the aerodynamic advantages of flying in close proxmity to the ground: Rostislav Evgenievich Alexeyev's Central Hydrofoil Design Bureau developed between the 1950s and early 1990s large aircraft with rectangular wings -- generally known as \textit{ekranoplan} -- that flew in extreme ground-effect with power-assisted take off \cite{Komissarov2010,Rozhdestvensky2006}. German aircraft engineer Alexander Lippisch took the route, in the 1960s, of reversed-delta wing designs -- generally known as \textit{Wing-in-ground-effect (WIG)} aircraft -- with a strong anhedral angle to constrain a cushion of air underneath the wing to enhance ground-effect \cite{Yun2010}.  

Recently, within the frame of electrical aviation and sustainability, interest in the development of wing-in-ground-effect craft has seen a renaissance in both unmanned and manned aviation with the development of the Regent Viceroy Seaglider \cite{Regent2024}. Other examples include Flying Ship Company cargo vehicle \cite{FSC2024}, and the Unmanned ground-effect Vehicle (UGEV) technology demonstrator \cite{Papadopoulos2022} designed at the Aristotle University of Thessaloniki. It is interesting to notice two different approaches in the development of these aircraft: whilst the Flying Ship Company relies on a classical two-element inverted delta wing with anhedral angle, Regent and researcher at the Aristotle University of Thessaloniki focused on novel wing designs for enhanced aerodynamic efficiency and multi-role flight capabilities. This is also the approach taken in ongoing work at KTH The Royal Institute of Technology \cite{Rejish2023}, of which the work presented here is a part. 

Of interest for the current study is the work completed at the Aristotle University of Thessaloniki \cite{Papadopoulos2022}, where different types of passive flow control have been implemented, including different designs of serrated leading-edges, a concept briefly explored also by Mehraban and Djavareshkian \cite{Mehraban2021} for an aircraft-type wing and by Arrondeau and Rana \cite{Arroundeau2020} for a multi-section wing for racing vehicles. Briefly highlighted, the majority of the work conducted on serrated leading-edges has been focused almost exclusively on wings, both of infinite and finite span, out-of-ground-effect at low Reynolds number. Results have consistently shown positive effects of the geometry on the flow separation characteristics, leading to an increase in achievable maximum lift coefficient and increase in stall angle of attack, coupled with a decrease in drag coefficient at those conditions \cite{Miklosovic2004,Favier2012,Lin2012}.  A correlation between wing planform and Reynolds number for serrated leading-edges has also been established by Miklosovic et al. \cite{Miklosovic2007}.

In summery, it was found that limited work has been conducted on the effects of serrated leading-edges for wings in-ground-effect and that even more limited work has been conducted on the effects of wing geometry variation. It should also be noted that the work by Papdopoulos et al. \cite{Papadopoulos2022} implemented these designs directly in the conceptual phase of an unmanned aerial vehicle, and Mehraban and Djavareshkian \cite{Mehraban2021} focused mainly on a single, constant-chord wing section.  

It is the objective of the current preliminary work to further investigate numerically and experimentally the effects of serrated - or tubercled - leading-edges at low Reynolds number on the aerodynamics of two-section, planar, wing in-ground-effect, using a more parametric approach by comparing two leading-edge geometries at two determined distances from the ground, and including effects of planform geometry.
\section{Experimental Set Up}\label{Sec2}
\subsection{Wind-Tunnel Facility}\label{Sec2.1}
\begin{figure}[t]
    \begin{center}
        \includegraphics[width=1.0\linewidth] {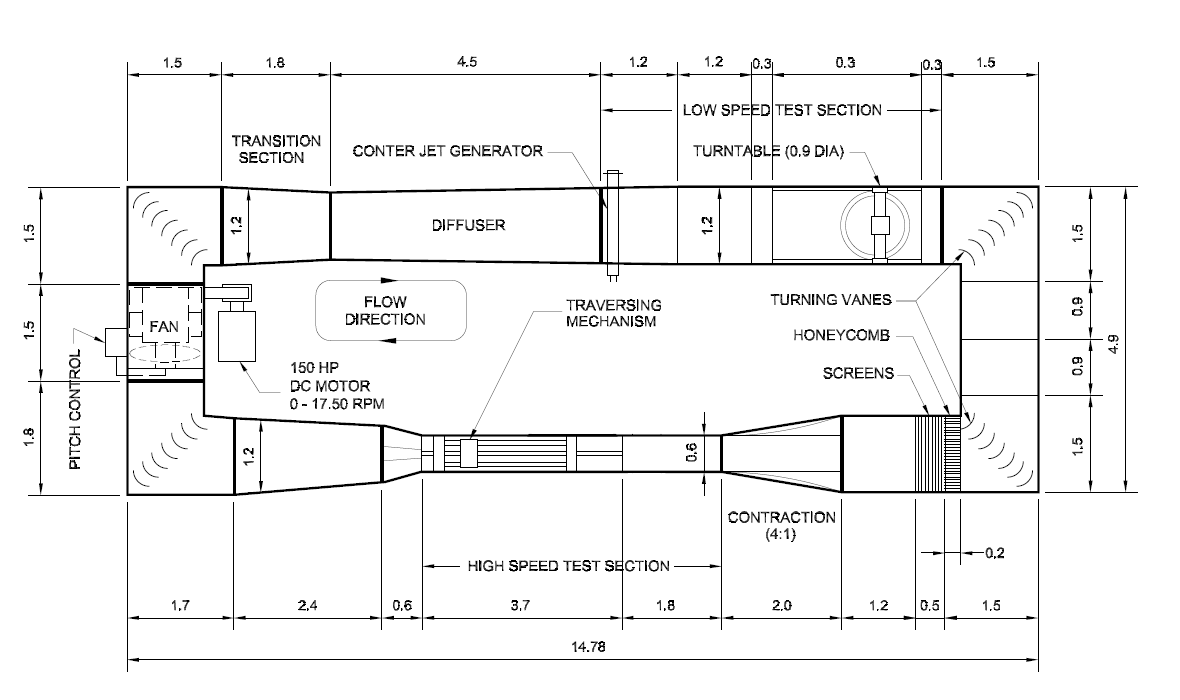}
    \end{center}        
    \caption{Schematic of the Morkovin Tunnel \cite{Gravante2003} - Adapted}
    \label{Fig:IIT_Tunnel}
\end{figure}

Tests were conducted in the Mark V. Morkovin atmospheric, closed-return wind-tunnel at Illinois Institute of Technology (IIT) shown in Figure  \ref{Fig:IIT_Tunnel}\cite{Gravante2003}. The wind-tunnel  has a low-speed and a high-speed test section, and the latter was used for the current experiments. The high-speed test section has a constant  cross-section of 0.61 m wide $\times$ 0.91 m high and 3.7 m long, and is preceded by a development section 1.8 m in length.

A smooth flat plate spanning the width of the test section, with the full length of the test section, and suspended 0.3 m above the tunnel floor was used to simulate the presence of the ground. The flat plate has an elliptical leading-edge and a transition strip placed 0.152 m downstream of the leading-edge which resulted in a fully developed boundary layer. 

\begin{figure}[h!]
    \begin{center}
        \includegraphics[width=1.0\linewidth]{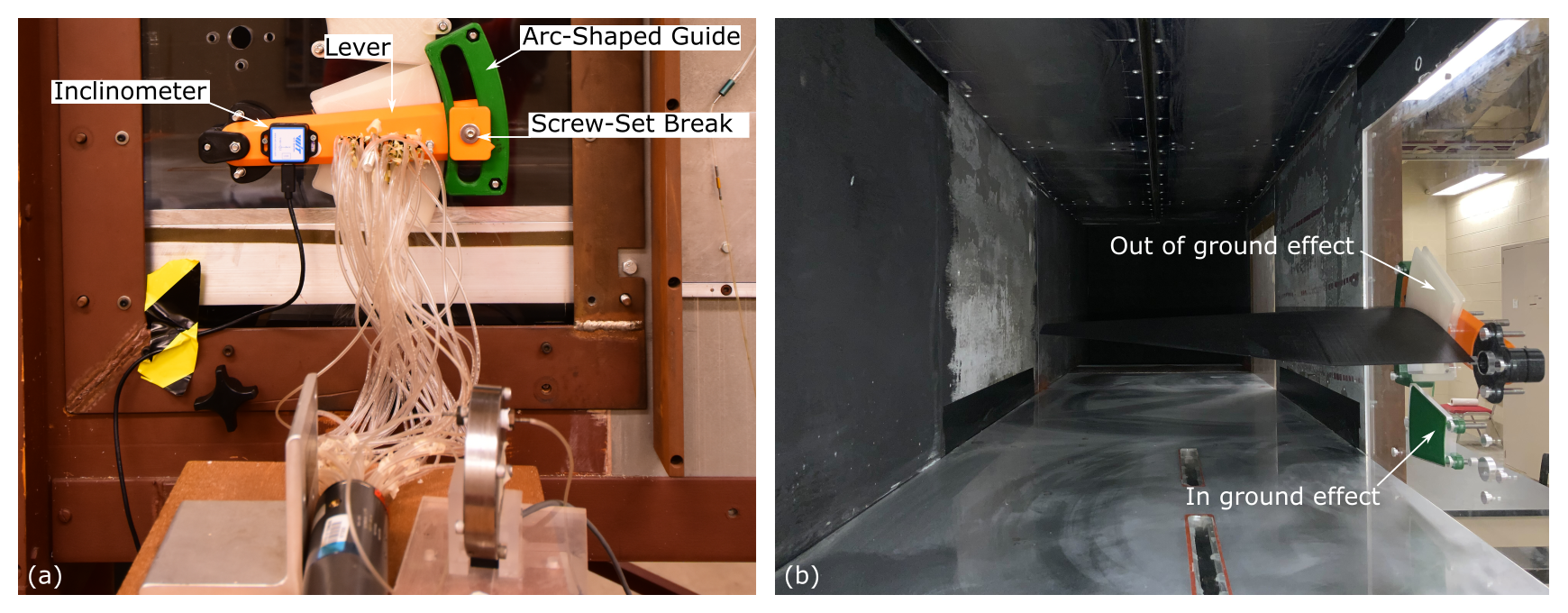}
    \end{center}        
    \caption{(a) Angle of Attack Mechanism; (b) Wing in the Test Section.}
    \label{Fig:WT_Set_Up}
\end{figure}

\subsection{Wind Tunnel Set Up}\label{Sec2.2}
The placement of the model was a compromise between the constraint imposed by the wind-tunnel existing set up and by an attempt at minimizing the impact of the boundary layer on the measurements, as the presence of the suspended plate boundary layer is not representative of an ideal testing condition and with a uniform upstream conditions.  The streamwise location of the model was selected not to allow the excessive growth of the suspended plate boundary layer.

The wind-tunnel was not equipped with a support for the current model, so a set up had to be designed and manufactured to allow the wing model to be mounted on the tunnel side wall. A three-part manual mechanism was implemented as shown in Figure \ref{Fig:WT_Set_Up}(a) consisting of a lever equipped with a rotation pin, onto which the wing was mounted and included an arc-shaped guide; and a set screw break to fix the model at selected angle of attack. A two-axis digital inclinometer was mounted to manually set the angle of attack of the wing model. To conform with convention, the set up was designed to allow model rotation about the trailing edge, to keep the height constant at the trailing-edge location as the angle of attack is varied. A provision was made in the set up with two sets of cuts in the window to allow for two testing ground heights. An example of the wing mounted in the test section is shown in Figure \ref{Fig:WT_Set_Up}(b).
\subsection{Model Design}\label{Sec2.3}
Two wing models were designed for the experiment: one with a straight leading-edge (W1), and one with a serrated leading-edge (W2). The wing is a two-element wing with an inboard section of constant chord and an outboard section with a trailing edge sweep angle of $\sim 26 deg$. The wing has has a constant NACA 63 312 section, and aspect ratio of 2.76 \cite{Rejish2023}. The airfoil was chosen as it has a flat loer surface advantageous for ground-effect flight. 

The wings for the experiment have a half-span of 0.43 m and a root chord and tip chord of 0.195 m and 0.05 m respectively, with a constant chord section with a span of 0.13 m and a section with a trailing-edge taper with a 0.3 m span. The models were manufactured using an Ultimaker3 3D printer using standard PLA material. As the model was larger than the printing volume, it had to be manufactured in three parts, requiring minor post-processing at assembly as a result of the printing tolerances and surface finishing at the joints.

\begin{figure}[t]
    \begin{center}
        \includegraphics[width=1.0\linewidth]{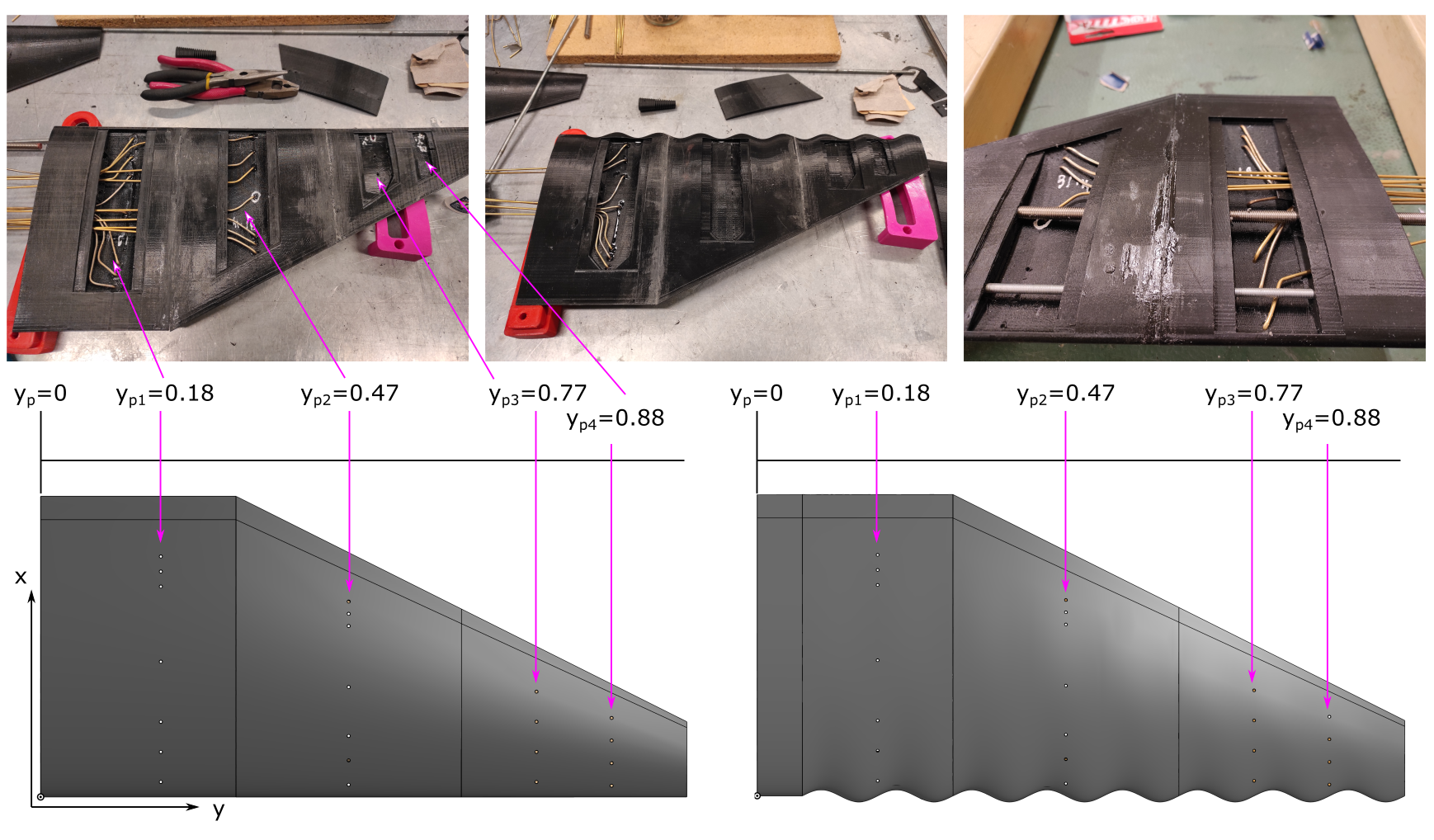}
    \end{center}        
    \caption{Instrumented Wings}
    \label{Fig:Wing}
\end{figure}

Two types of leading-edges were designed, a classical straight leading-edge (Figure \ref{Fig:Wing} left) and a serrated leading-edge (Figure \ref{Fig:Wing} right), and each model was instrumented with four section of pressure taps, with the spanwise location of the taps kept consistent between the two model. No serration is present at the trailing-edge, which retains a straight line geometry.

Chordwise  and spanwise locations of the pressure taps are shown graphically in Figure \ref{Fig:Wing}, and their coordinates are presented in Table \ref{tab:tap_location}. The number of chordwise pressure taps per-section varied with chord length due to internal space constraints. Pressure tap locations are shown as fractions of the chord ($x_{u_n}/c;x_{l_n}/c$) and of the wing semi-span ($y_{p_n}$). The subscripts $u$ and $l$ correspond to pressure taps on the upper and lower wing surface, respectively, and the numerical subscript indicates the location from the leading edge or form the wing origin. 

\begin{table}[h!]
\centering
\begin{tabular}{l c c c c c c r}
  \hline
  & $x_{u_1}/c$  & $x_{u_2}/c$ & $x_{u_3}/c$ & $x_{u_4}/c$ & $x_{u_5}/c$ & $x_{u_6}/c$ & $x_{u_7}/c$ \\
  \hline
    $y_{p_1}=0.18$ & 0.05 & 0.15 & 0.25 & 0.45 & 0.70 & 0.75 & 0.80  \\ 
    $y_{p_2}=0.47$ & 0.05 & 0.15 & 0.25 & 0.45 & 0.70 & 0.75 & 0.80  \\ 
    $y_{p_3}=0.76$ & 0.10 & 0.30 & 0.50 & 0.70 &      &      &       \\ 
    $y_{p_4}=0.88$ & 0.10 & 0.30 & 0.50 & 0.70 &      &      &       \\
    \hline
    \hline
     & $x_{l_1}/c$  & $x_{l_2}/c$ & $x_{l_3}/c$ & $x_{l_4}/c$ & $x_{l_5}/c$ & $x_{l_6}/c$ & $x_{l_7}/c$ \\
  \hline
    $y_{p_1}=0.18$ & 0.05 & 0.15 & 0.25 & 0.45 & 0.70 & 0.75 & 0.80 \\ 
    $y_{p_2}=0.47$ & 0.05 & 0.15 & 0.25 & 0.45 & 0.70 & 0.75 & 0.80 \\ 
    $y_{p_3}=0.76$ & 0.12 & 0.33 & 0.49 & 0.69 &      &      &      \\ 
    $y_{p_4}=0.88$ & 0.13 & 0.33 & 0.47 & 0.67 &      &      &      \\
    \hline
    \end{tabular}
    \caption{Chordwise and spanwise pressure taps locations.}
    \label{tab:tap_location}
\end{table}

Pressure taps had an internal diameter of 1.6 mm and were connected to the outside of the model using brass tubing to securely route the air lines and avoid potential pinching often present when using soft tygon tubing. The instrumented wings are shown in Figure \ref{Fig:Wing}. 
\subsection{Acquisition System}\label{Sec2.4}
\begin{figure}[h!]
    \begin{center}
        \includegraphics[width=1.0\linewidth]{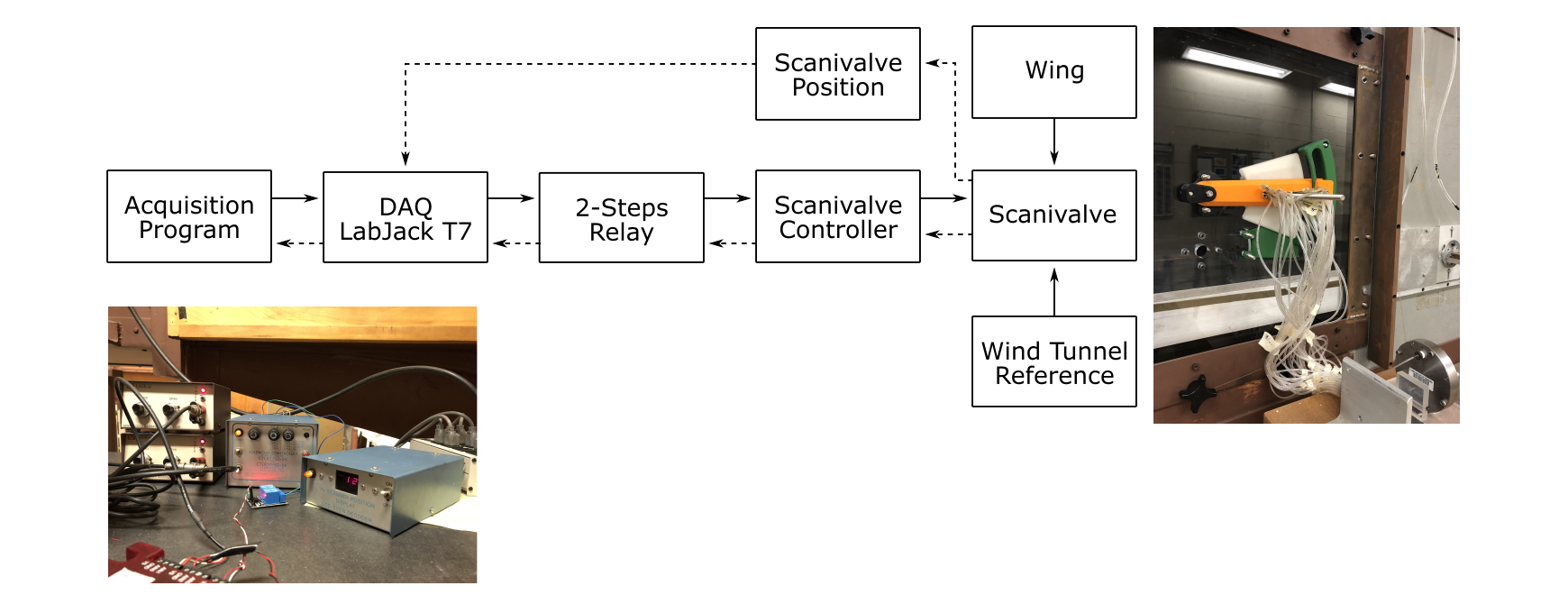}
    \end{center}        
    \caption{Acquisition System Diagram}
    \label{Fig:DAQ}
\end{figure}

Each wing was instrumented with 48 pressure taps, 44 for the wing and four for the wind tunnel reference system, connected to a rotating multi-port pressure Scanivalve Solenoid 48J9-2635. The output of the Scanivalve was measured with a Validyne Model DP103-26 differential transducer. Data is acquired through a LabJack T7 DAQ connected to a two-step relay module which is in turn connected to the Scanivalve to advance its steps. A diagram and images of the system are shown in Figure \ref{Fig:DAQ}. The acquisition rate was set to $\sim 130$ Hz with a 3 sec acquisition time providing 400 samples.   

\section{Numerical Set Up}\label{Sec3}
\subsection{Domain Geometry}
The numerical analysis was performed in ANSYS Fluent using Reynolds-averaged Navier-Stokes (RANS) simulations for  the straight and serrated leading-edge wings, considering both in- and out of ground-effect \cite{Vinuesa2018}.

Fluid-domain dimensions were chosen to minimize the impact of the boundary conditions on the solution \cite{john1995computational}, with special considerations for ground-effects simulations. A box-type computational domain was implemented, specific to each wing position, out of- or in-ground-effect, and for leading-edge geometry.

A computational domain was implemented for the wing with the straight leading-edge consisting of of 13 cord-lengths \textit{c}, where \textit{c} represents the mean aerodynamic chord in the streamwise direction (3c upstream and 9c downstream of the wing), 6c in the spanwise direction with origin at the root chord, and 10c in the vertical direction evenly distributed with 5c above and 5c below the wing. The streamwise length of the domain was increased for the wing with the serrated leading-edge in consideration of the expected vortical flow formation, with an upstream length of 10c and a downstream length of 25c to allow for longer wake dissipation. Boundaries were set with a Velocity Inlet, Pressure Outlet, Symmetry Plane at the wing root, and No-Slip Walls on the far side and above and below the wing.

The computational domain for in-ground-effect simulations was modified by replacing the No-Slip Wall boundary with a Slip-Wall boundary below the wing to correctly simulate the ground. The wings were placed at a height of 0.4c -- with reference to the trailing-edge position -- from the ground to represent the in-ground-effect conditions.   
\subsection{Flow Conditions}
The free stream of the tunnel is known to have 0.1$\%$ turbulence intensity and the free stream velocity was maintained at $17.8$ m/s throughout the tests.  The test  section pressure and density conditions are nearly the same as the laboratory atmospheric conditions. The flow conditions were selected to closely represent the average chord Reynolds number of $\sim 200,000$. 

Two different approaches were used when setting the flow velocity conditions at varying angles of attack for numerical simulations. For the case of the wing out-of-ground-effect the variation in angle of attack was simulated by setting the vertical and horizontal components of the free-stream velocity as inputs and keeping the mesh and geometry unchanged. For the wing in-ground-effect, due to the presence of the ground, angle of attack variations were obtained by geometrically rotate the wing about its trailing-edge, with velocity conditions kept aligned with the domain horizontal axis.  
\subsection{Mesh Geometry and Validation}
Polyhedral cells were implemented for all configurations. In near-wall regions, where a boundary layer is expected to form, semi-structured prismatic cells were employed, keeping the $y^+<1$ by adjusting the height of the first layer at the wall, with a denser mesh applied near the wing with cell inflation increasing far from the wing.

\begin{figure}[t]
    \begin{center}
    \subfigure[]{\includegraphics[width=0.45\linewidth]{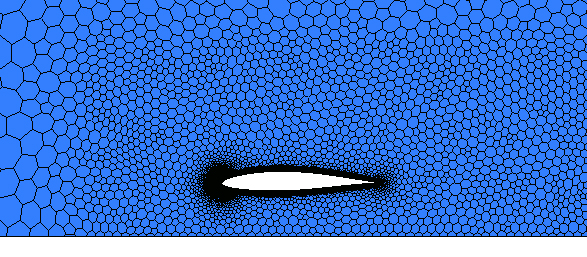}}
    \subfigure[]{\includegraphics[width=0.45\linewidth]{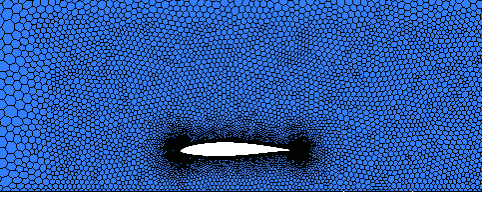}}
    \end{center}
    \caption{Representative meshes for the Straight Wing (a) and the Serrated Wing (b) in ground-effect}
    \label{Meshes_Comparison}
\end{figure}

Additional surface and volumetric refinements were applied to obtain a sufficiently fine grid near the wing, with a particular focus to the leading-edge, the trailing-edge, the tip of the wing and the wake region. For the wing with the serrated leading-edge, a finer mesh was employed to properly discretize the sinusoidal geometry and to more accurately capture the expected counter-rotating vortices at the leading-edge. Examples of the mesh for the in-ground-effect set up are shown in Figure \ref{Meshes_Comparison}. 

A mesh independence analysis was conducted for both wing leading-edge geometries and for both out-of- and in-ground-effect, as the complexity of the flow is expected to vary among these fours numerical configurations. The variation of lift and drag coefficients was evaluated among different mesh densities, resulting in a mesh with a final cell-count of 2,670,014 for the straight wing and 6,599,244 for the serrated wing out of ground-effect, and to 2,757,385 for the straight wing and 5,829,956 for the serrated wing in ground-effect. These represented a variation of $0.005\%$ for the lift coefficient and $0.03\%$ between the chosen mesh and the one with the highest refinement.   

Different turbulence models were tested to asses accuracy compared to experimental data. The one-equation Spalart-Allmaras \cite{spalart1992one} and the two-equation SST $k-\omega$ \cite{Menter1994} models were compared, and the Spalart-Allmaras model was selected due to its computational efficiency and acceptable accuracy \cite{Vinuesa2014}. The low speed of the free stream led to choose a pressure-based solver, suitable for incompressible flows, and the pressure-velocity coupling was obtained thanks to the SIMPLE algorithm. For angles of attack up to 10 degrees, the second-order accuracy was used for both pressure and momentum equations. A lower-fidelity model was implemented at higher angles of attack. 
\section{Discussion of Results}\label{Sec4}
Numerical results for four cases, corresponding to the straight leading-edge wing (W1) and the serrated leading-edge wing (W2) out of ground-effect and in ground-effect, with a ground clearance of $H = 0.4c_{MAC}$ are presented, where $c_{MAC}$ is the mean aerodynamic chord of the wing.
\begin{figure}[t]
    \begin{center}
    \subfigure[]{\includegraphics[width=0.45\linewidth]{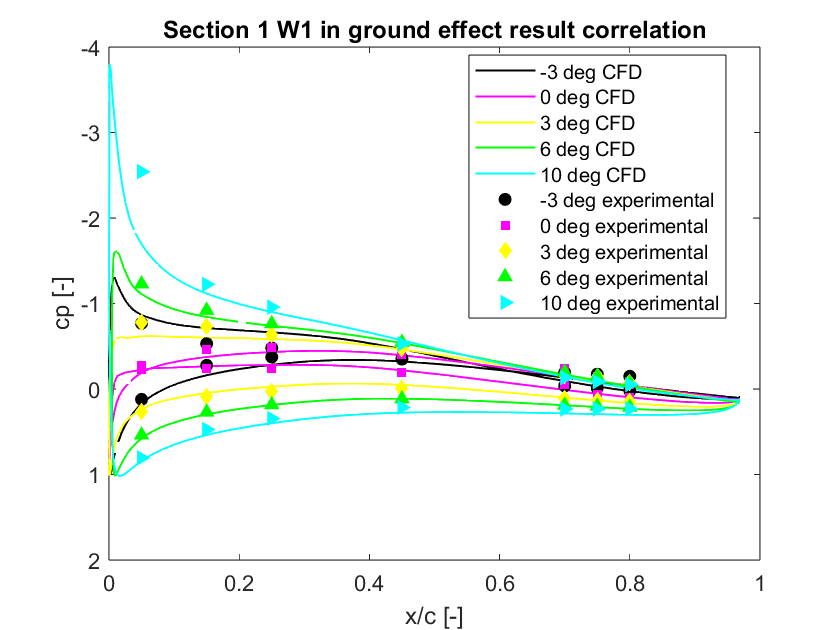}}
    \subfigure[]{\includegraphics[width=0.45\linewidth]{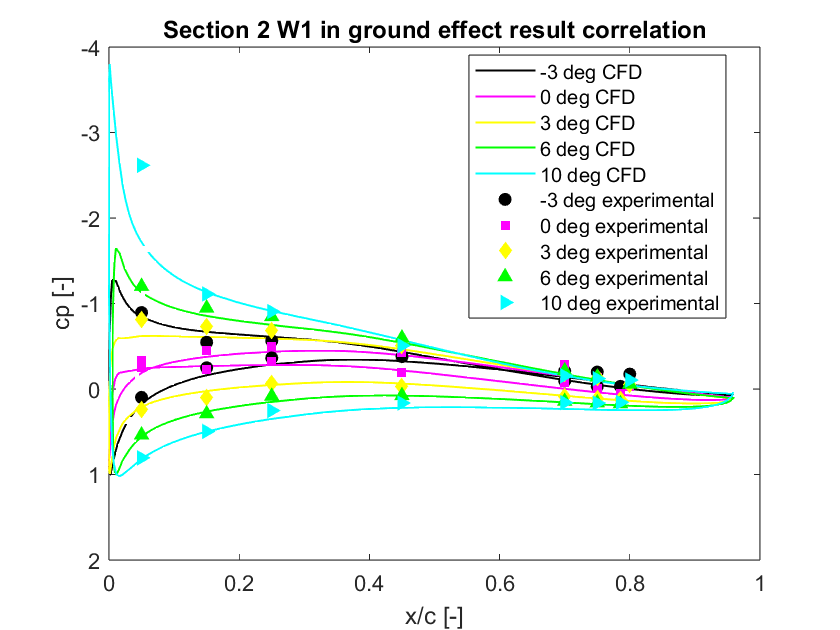}}
    \subfigure[]{\includegraphics[width=0.45\linewidth]{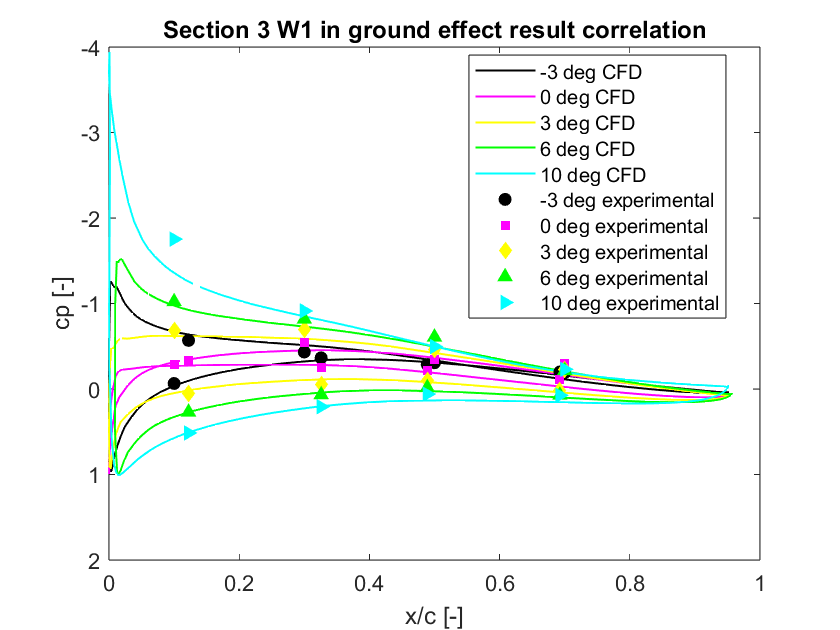}}
    \subfigure[]{\includegraphics[width=0.45\linewidth]{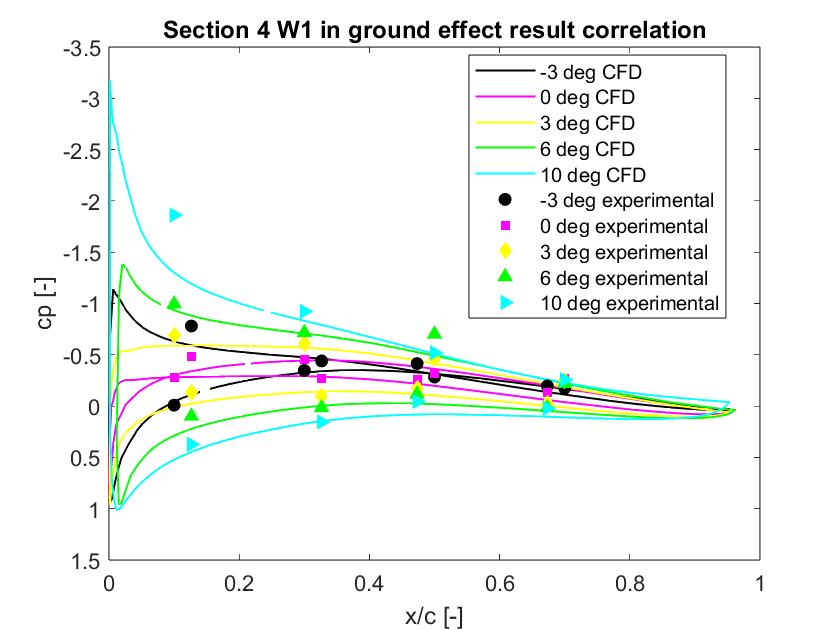}}
    \end{center}
    \caption{Correlation of experimental and numerical results for the straight leading-edge wing in ground-effect for sections (a) $y_{p_1}=0.18$, (b) $y_{p_1}=0.47$, (c) $y_{p_1}=0.76$, and (d) $y_{p_1}=0.88$.}
    \label{W1_IN_Datacorrelation}
\end{figure}
\subsection{Chordwise Pressure Coefficient Profiles - Validation of Numerical Data}\label{ResultsCp}
Pressure coefficient data obtained from the experiment were used for the validation of the results produced with numerical simulations. Experimental data were collected for a range of angles of attack between -4 deg and 15 deg. 

\begin{figure}[t]
    \begin{center}
    \subfigure[]{\includegraphics[width=0.45\linewidth]{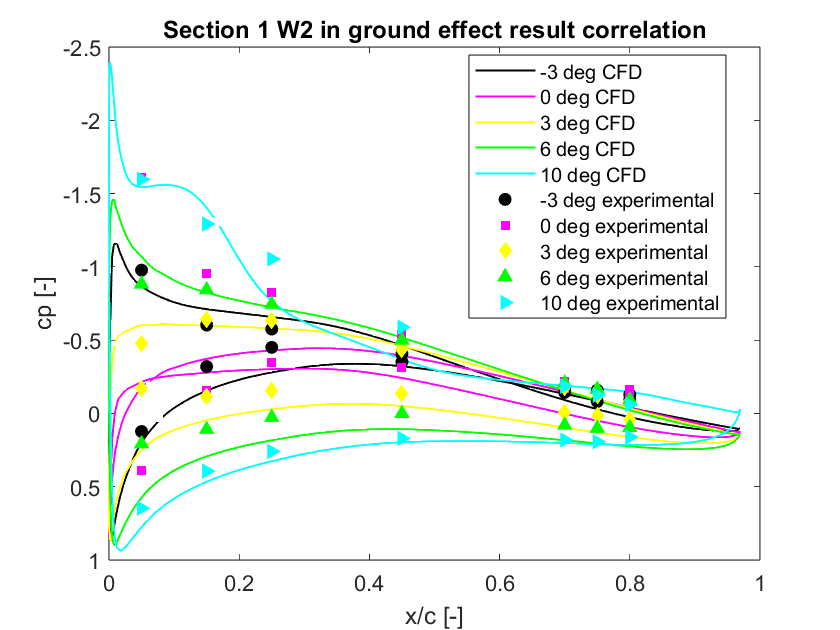}}
    \subfigure[]{\includegraphics[width=0.45\linewidth]{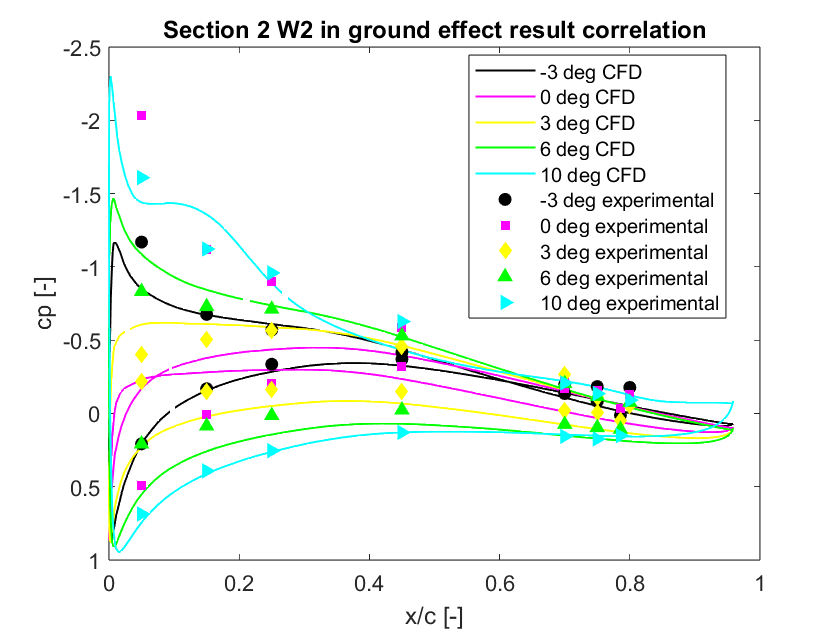}}
    \subfigure[]{\includegraphics[width=0.45\linewidth]{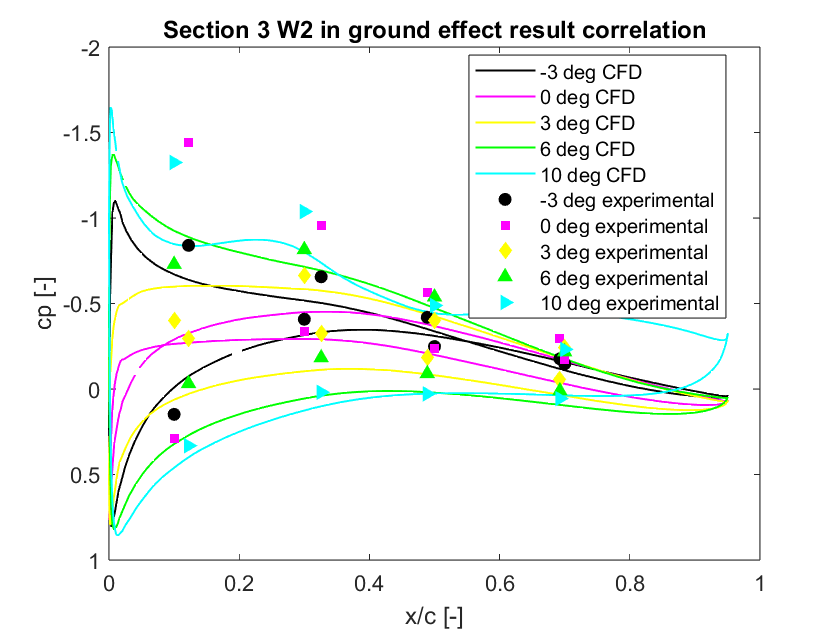}}
    \subfigure[]{\includegraphics[width=0.45\linewidth]{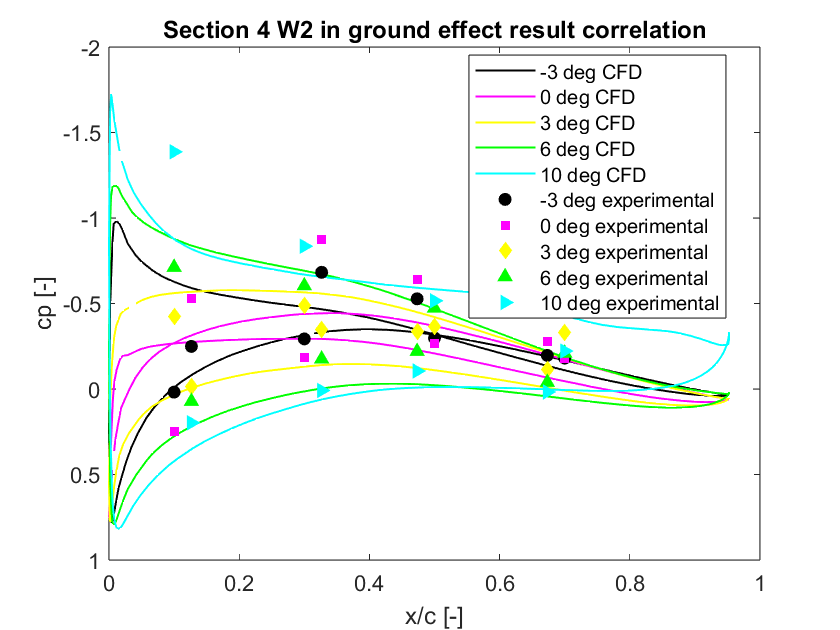}}
    \end{center}
    \caption{Correlation of experimental and numerical results for the serrated leading-edge wing in ground-effect for sections (a) $y_{p_1}=0.18$, (b) $y_{p_1}=0.47$, (c) $y_{p_1}=0.76$, and (d) $y_{p_1}=0.88$.}
    \label{W2_IN_Datacorrelation}
\end{figure}

Samples of the comparison of numerical and experimental data for the wings in ground effect are shown in Figure \ref{W1_IN_Datacorrelation} and Figure \ref{W2_IN_Datacorrelation} for the wing with the straight and with the serrated leading-edge, respectively. These results are the more interesting as they present a more complex flow and therefore provide better support for the validation process of numerical data. 

Overall, good agreement is shown, although numerical results appear to consistently underpredict pressure distributions for the serrated leading-edge wing, including for the suction peak strength. Near the suction peak the acceleration of the flow is the highest and this area of the flow is shown to be the most difficult to match between experimental and numerical data. Again, this is observed most noticeably for the wing with the serrated leading-edge geometry. Two factors may be the source of these discrepancies: (a) the pressure port $x_{u_{1}}$ was placed close to the location of the vortex generation, leading to unsteadiness in the flow which required longer acquisition time for an accurate averaged value; (b) numerical set up at the leading-edge for the numerical study. Hence, the differences between the results appear to be attributable to a combination of numerical and experimental uncertainty.

Furthermore, the underprediction of pressure becomes more significant for wing sections closer to the wing tip and as the angle of attack $\alpha$ is increased. This may be explained by the inability of the current numerical set up to correctly capture the increasing vortex strength and the interaction of the vortices generated at the leading-edge with wing tip vortices. This is shown in Figure \ref{W2_IN_Datacorrelation}(d), where the maximum disagreement of $\sim 37\%$ in $C_p$ at the region of the suction peak is present. Finally, numerical results show a much stronger separation at the trailing edge for the wing with serrated leading-edges compared to experimental data, particularly towards the outboard section where trailing-edge sweep is present.

Overall, as these results are exploratory in nature, it is estimated that numerical data is reasonably validated within the limitations of both experimental and numerical set ups.

\subsection{Aerodynamic Analysis}\label{ResultsAero}
\begin{figure}[t]
    \centering
    \subfigure[]{\includegraphics[width=0.45\linewidth]{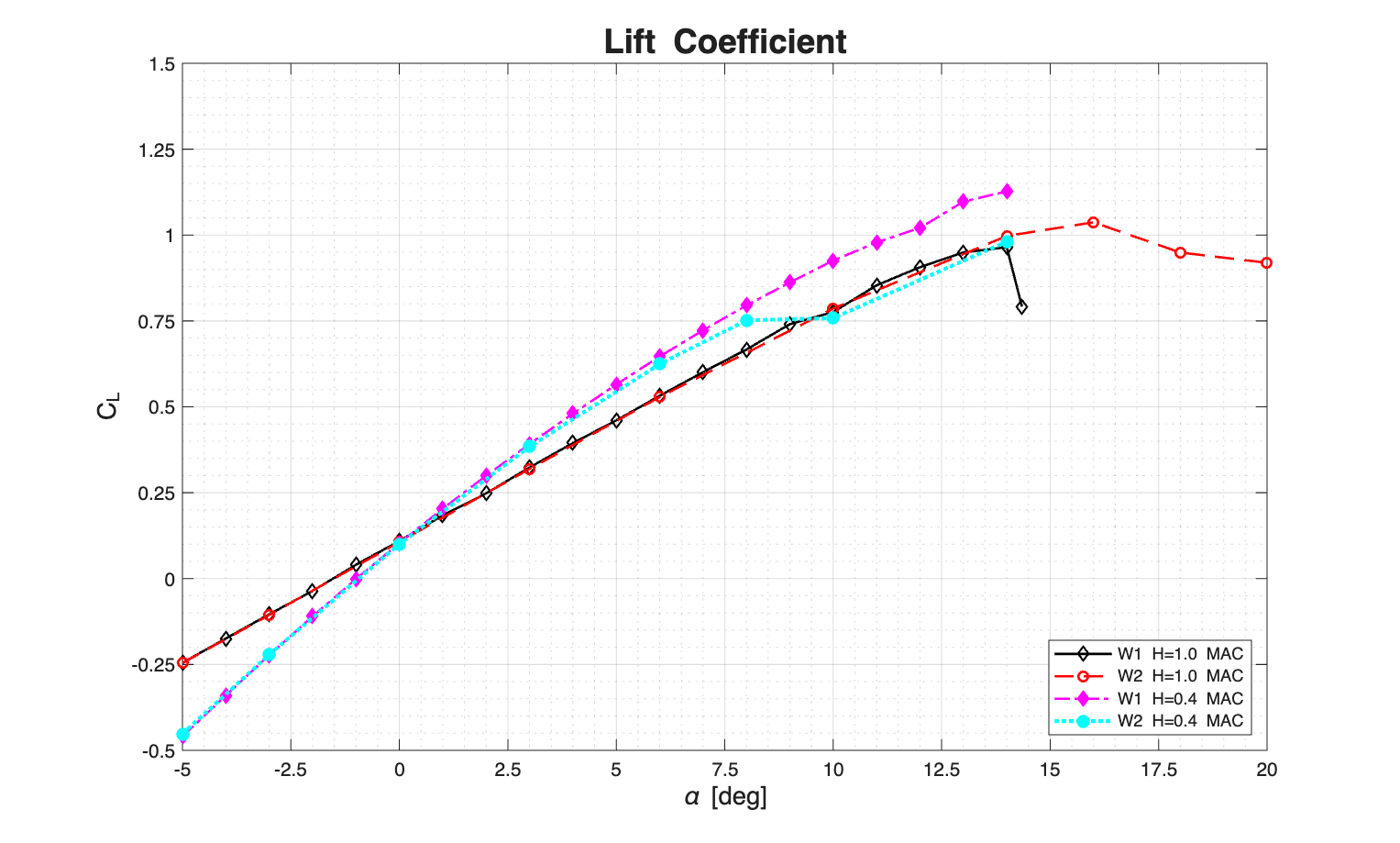}} 
    \subfigure[]{\includegraphics[width=0.45\linewidth]{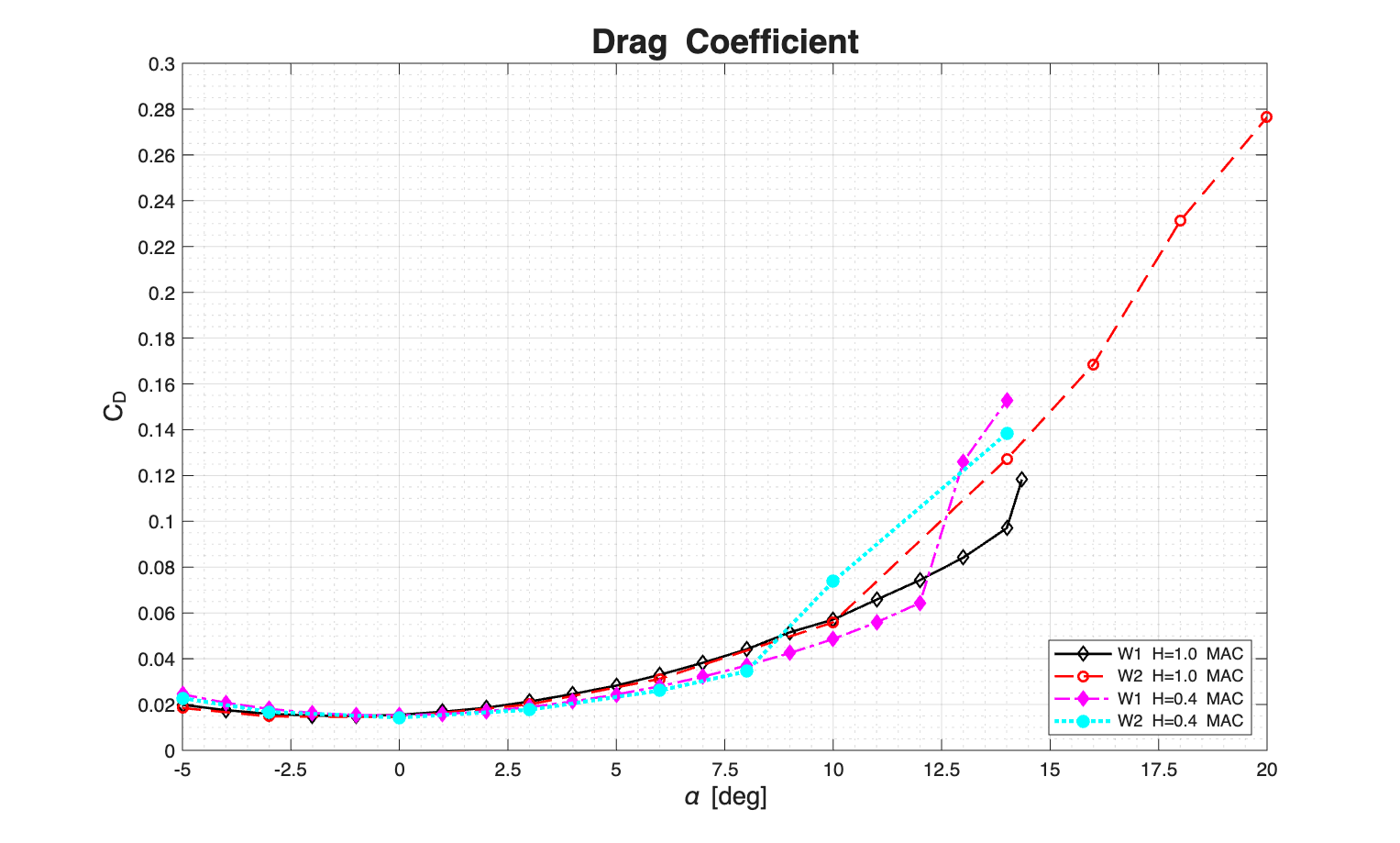}}
    \subfigure[]{\includegraphics[width=0.45\linewidth]{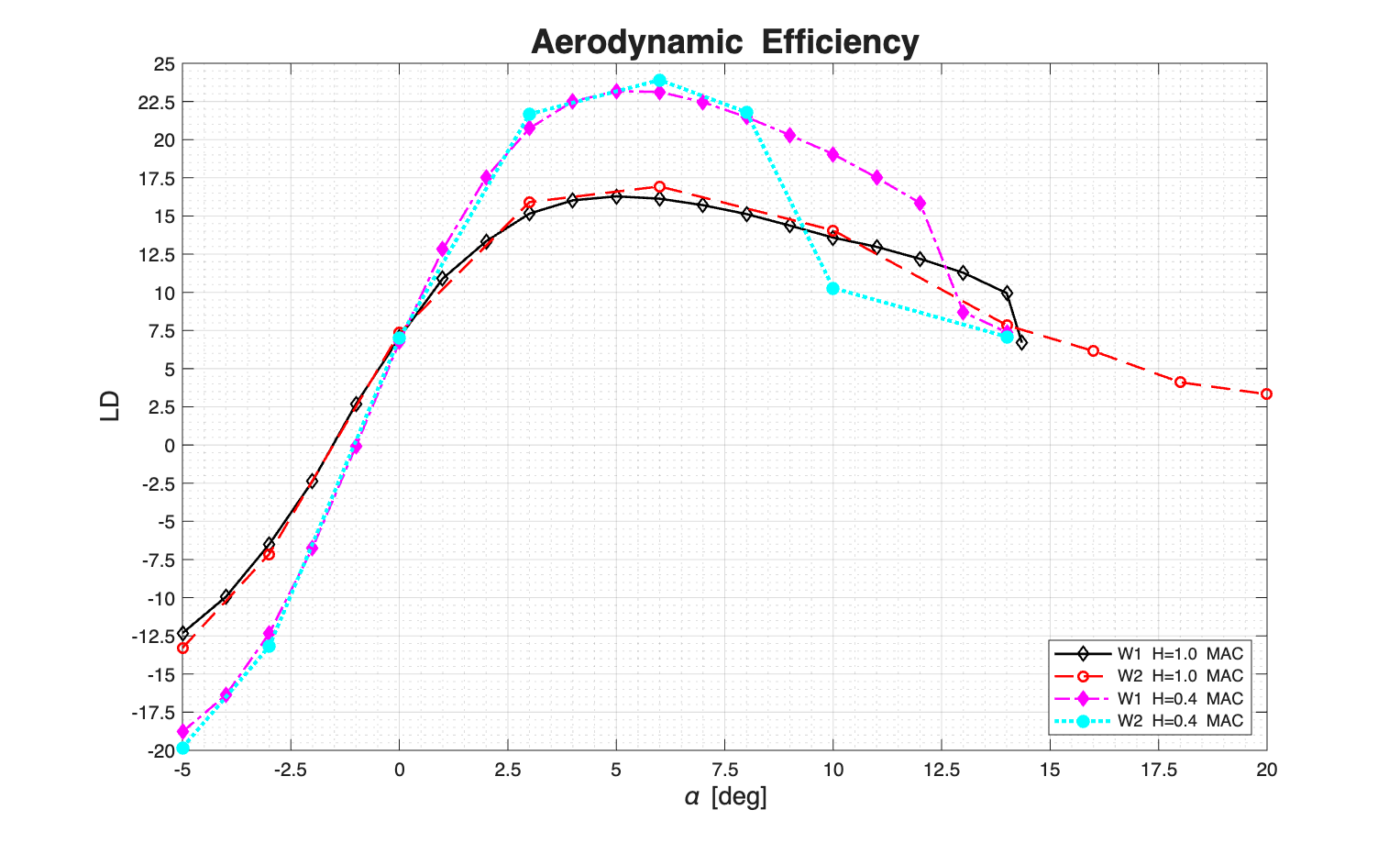}}
    \subfigure[]{\includegraphics[width=0.45\linewidth]{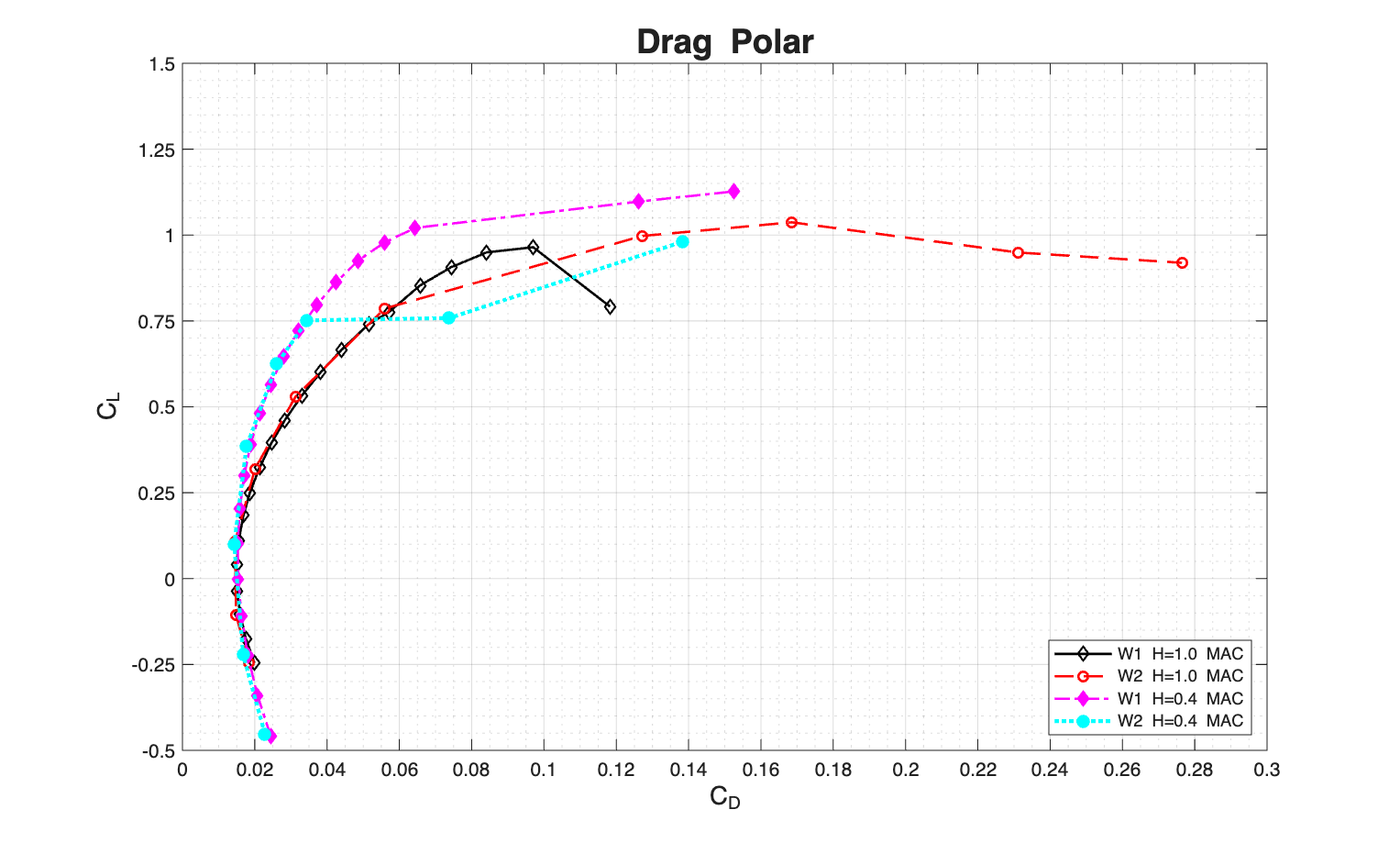}}
    \caption{Aerodynamic performance of wing W1 and W2 (a) Lift curve, (b) Drag Curve and (c) Aerodynamic Efficiency (d) Drag Polar} 
    \label{aero_coefficients_graphs}
\end{figure}

Figure \ref{aero_coefficients_graphs}(a) shows that the wing has a ``leading-edge type´´ stall, as indicated by the sudden drop in lift coefficient at $\alpha \sim 14 deg$, which is expected has it has a relatively small leading-edge radius conducive to this type of stall despite a thickness ratio of $12\%$.

The primary effect for a wing in free-flow, as shown in literature \cite{Fish1995, Miklosovic2007} of the modification of a wing leading-edge from a straight geometry to a serrated one is to increase the maximum lift coefficient $C_{L_{max}}$ and stall angle $\alpha_{stall}$ of a wing. 
 
Current results, as shown in Figure \ref{aero_coefficients_graphs}(a) for the wings out-of-ground-effect, demonstrate that higher values for $C_{L_{max}}$ of $\sim 7.5\%$ and $\alpha_{stall}$ are achieved by the wing with the serrated leading-edge, compared to the one with the straight leading-edge. Results also show a change in the type of stall, from a sudden ``leading-edge type´´to a smoother ``trailing-edge type´´of stall. This is achieved by the generation of counter-rotating vortices at each serration trough at chordwise locations in the proximity of the pressure peak of the wing. The serration at the leading edge promotes vortex stability and higher vortex strength, similar to the effects of classical vortex generators \cite{Mehraban2021,Neves2024}, delaying flow separation. These results are in agreement with existing literature.   

\begin{figure}[hb]
    \begin{center}
        \includegraphics[width=1.0\linewidth]{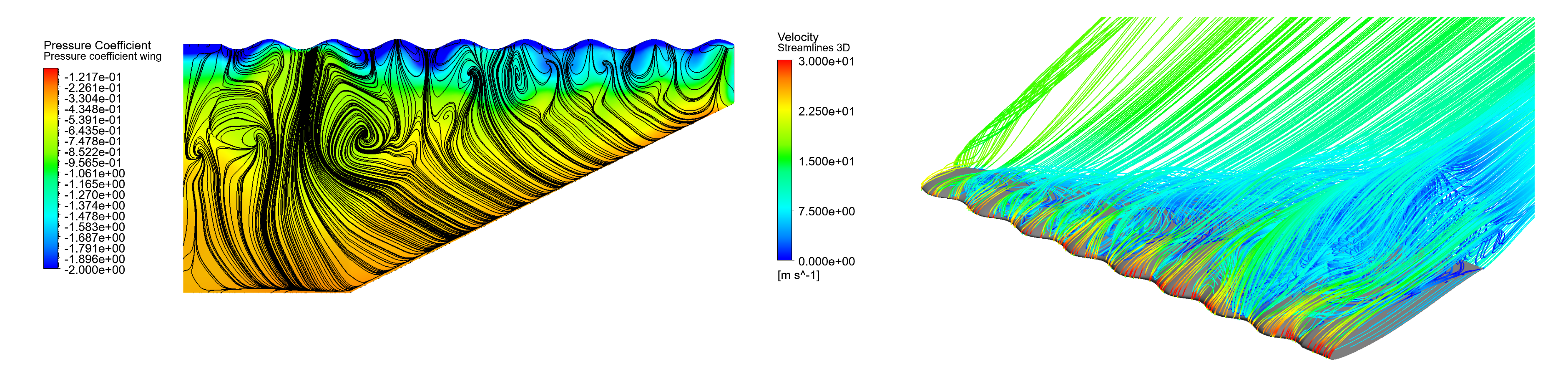}
    \end{center}        
    \caption{Surface pressure coefficient contour and shear stress lines over the upper surface of the tubercled leading-edge wing (left) out of ground-effect and numerical streamline flow visualization of post-stall flow at 18deg angle of attack out of ground-effect.}
    \label{streamlines_W2_18deg}
\end{figure}

The strong effects induced by the presence of the counter-rotating vortex pairs in delaying stall can be visually appreciated in Figure \ref{streamlines_W2_18deg} for wing W2 at $\alpha=18 deg$.

\begin{figure}
    \centering
    \includegraphics[width=0.9\linewidth]{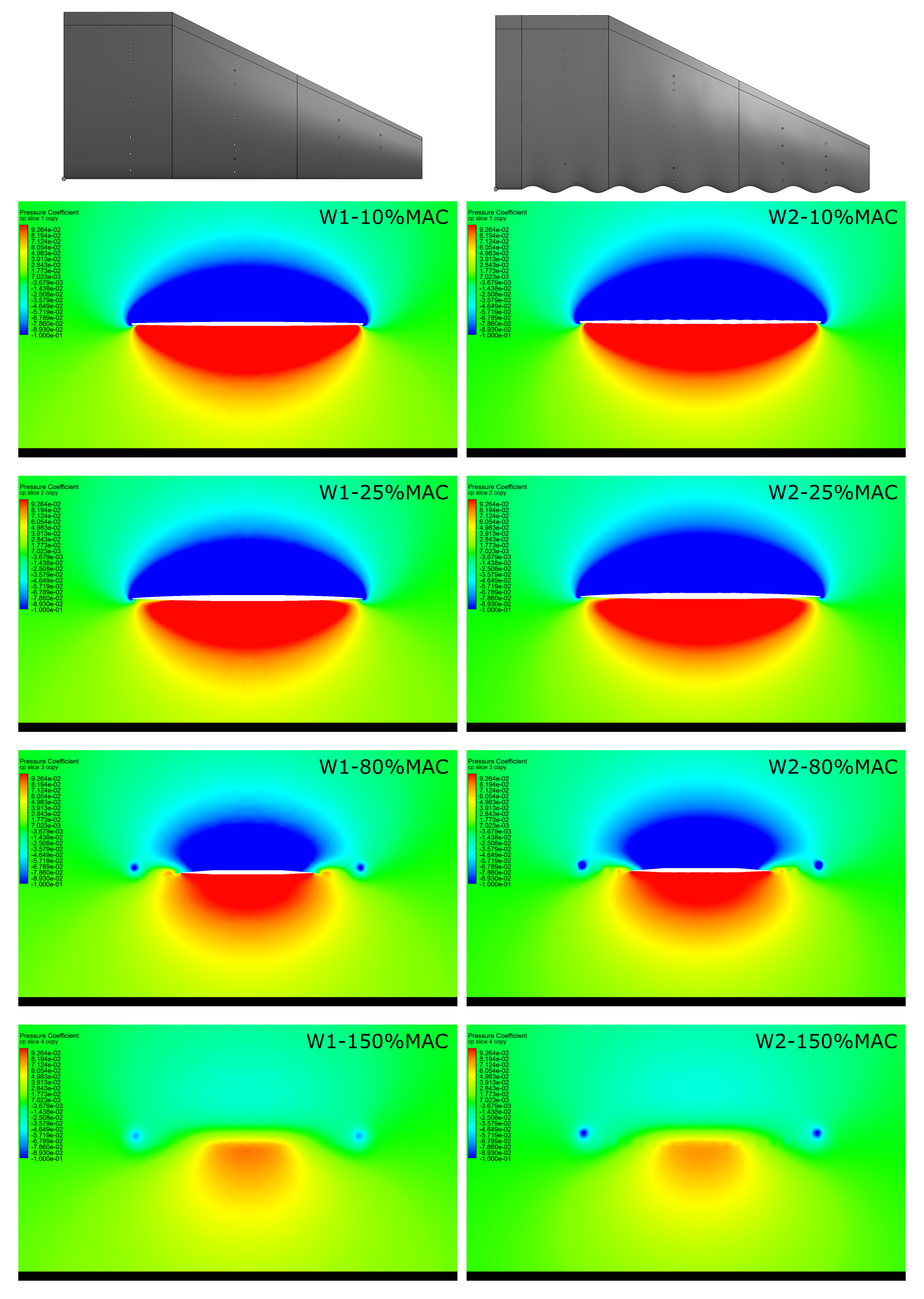}
    \caption{Spanwise flow field pressure distribution for W1 and W2 out of ground-effect}
    \label{Fig:spanwise_Cp_oge}
\end{figure}

Figure \ref{streamlines_W2_18deg} shows a well defined suction peak at the leading edge of the wing at an angle of attack when a wing with this geometry and cross section would normally be fully stalled, explaining the transition to a gentler, ``trailing-edge type´´ stall for the wing. Strong cross flow is clearly visible, and which mimics the expected flow at lower angles of attack for a wing with a straight leading-edge.   

Figure \ref{Fig:spanwise_Cp_oge} (mostly qualitatively) at $\alpha = 0 deg$ shows a decrease in high-pressure over the lower surface of the wing as an effect of the vortices generated by the serrated leading-edge, and stronger wing tip vortices that require more time to dissipate, as shown at location $150\%$ of MAC. The increase in strength of the wing tip vortices may be a consequence of the mixing of the wing tip vortex with vortex pair generated by the most-outboard serration.

As discussed in Section \ref{Sec1} -- Introduction, there is limited work on the effects of wings of finite span with serrated leading-edgesc and/or of the wing in-ground-effect, and there appears to be no available work on the effects of planform geometry. It is therefore of more relevance to focus the discussion on the effects of serration on the leading edge on a two-element wing in ground effect, and compare with the case of a wing of equivalent planform geometry and area with a straight leading-edge.

\begin{figure}[t]
    \centering
    \includegraphics[width=0.9\linewidth]{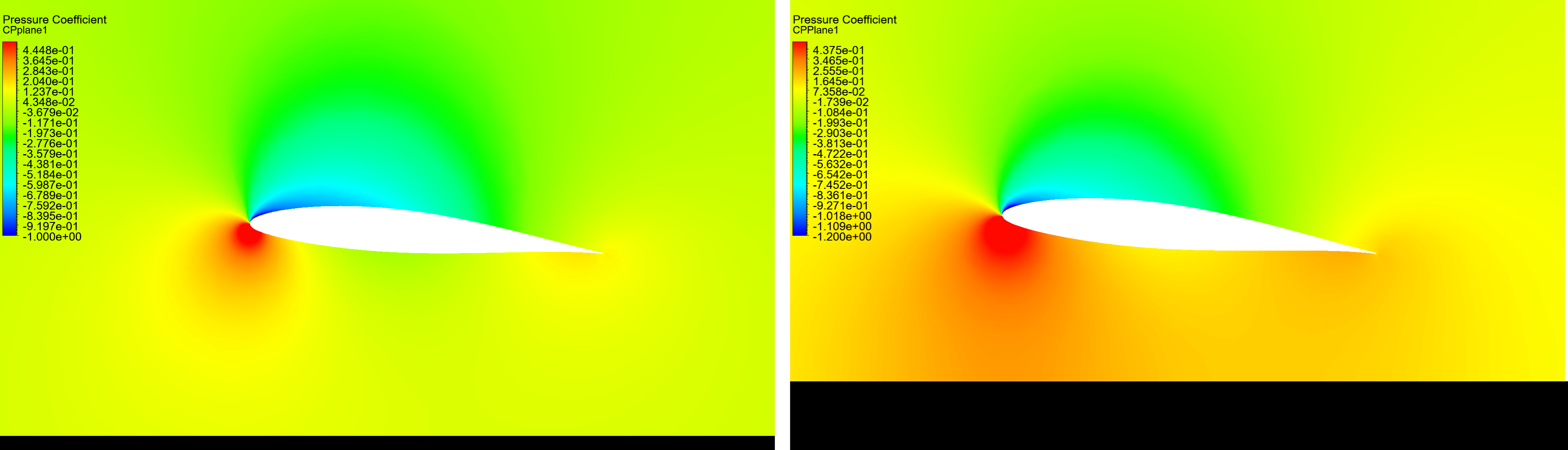}
    \caption{Pressure coefficient contour at $y_{p_{1}}$ of W1 out-of- and in-ground effect}
    \label{cp-profiles}
\end{figure}

Figure \ref{aero_coefficients_graphs}(b) and Figure \ref{aero_coefficients_graphs}(c) shows an overall improvement of the performance of the wings as they are subjected to ground effect, with a decrease in $C_D$ and an increase in aerodynamic efficiency at low angles of attack, as it is expected since wing tip vortices are bound by the presence of the ground, downwash is constrained, and the static pressure on the lower surface of the wing is increased as a consequence of the proximity of the ground \cite{Gudmunson2022}, as shown in Figure \ref{cp-profiles}.

The assessment of the performance of the W2 with the serrated leading-edge is more complex, in particular at high values of angle of attack, due to the difficulties in accurately capture the vortices being generated at the trough of the serrations. At low angles of attack, Figure \ref{aero_coefficients_graphs}(c) shows a small improvement in the range of $\sim 3.3\%$ in terms of aerodynamic efficiency, demonstrating that the presence of the vortices has a positive effect on the overall performance of the wing, with apparent negligibles penalties in drag, as shown in Figure \ref{aero_coefficients_graphs}(b).  

\begin{figure}
    \centering
    \includegraphics[width=0.9\linewidth]{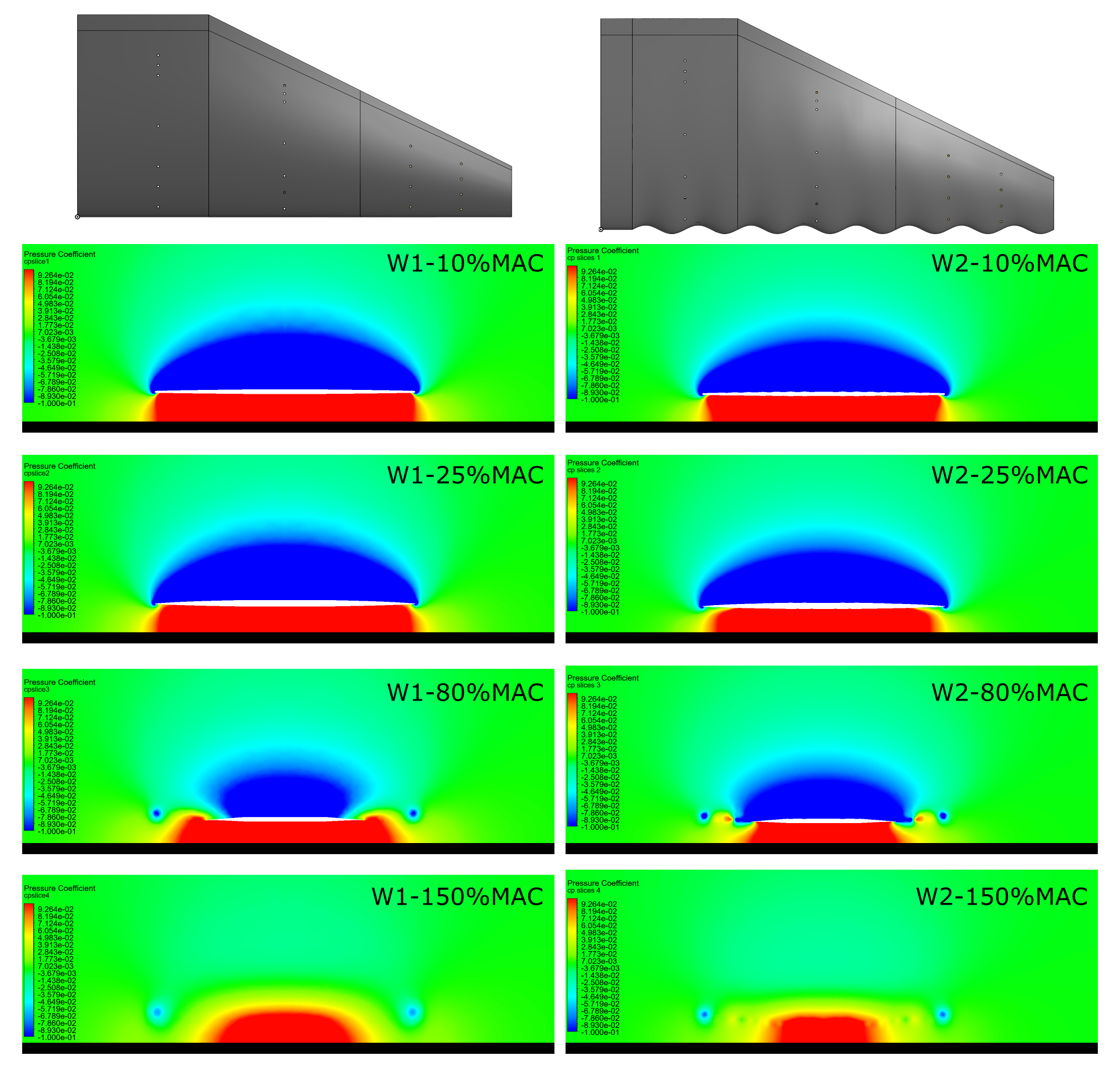}
    \caption{Spanwise flow field pressure distribution for W1 and W2 in of ground-effect}
    \label{Fig:spanwise_Cp_ige}
\end{figure}

Serrations at the leading-edge appear to have a much greater impact on pressure distribution -- both on the suction and pressure side -- compared to the case of the wings out-of-ground-effect. Figure \ref{Fig:spanwise_Cp_ige} shows that, at ``early streamwise locations, such as $10\% MAC$ and $25\% MAC$, the pressure distributions on the upper and lower surfaces are comparable, although with a shorter vertical extension from the upper surface, and with a narrower high-pressure region spanwise. This is most visible at $25\% MAC$ where the low-pressure region on the lower surface extends farther inboard at the wing tip. As pressure fields planes are evaluated farther downstream the differences between the two pressure fields become more marked. At $80\% MAC$ a high-pressure region is visible for the wing with the straight leading-edge ``spills-out´´ to generate a wider region of high static pressure, while for the wing with the serrated leading-edge the high-pressure region is confined to the actual wing-span. Oppositely, a single vortex is visible at $80\% MAC$ for the wing with the straight leading-edge stemming from the wing tip, as expected, while for the wing with serrated leading-edge a pair of counter-rotating vortices are clearly visible: one stemming from the wing tip and one extending from the serration at the outboard section of the wing. As a result of the complex system of counter-rotating vortices generated by the serrations at the leading-edge and the displacement of the momentum thickness, the region of low-pressure over the upper surface of the wing is extended spanwise, compensating the local loss of high-static-spanwise pressure distribution (compared to the wing with the straight leading edge) and overall still providing a marginal improvement to the overall performance of the wing at low angles of attack.   

\begin{figure}[t]
    \centering
    {\includegraphics[width=1.0\linewidth]{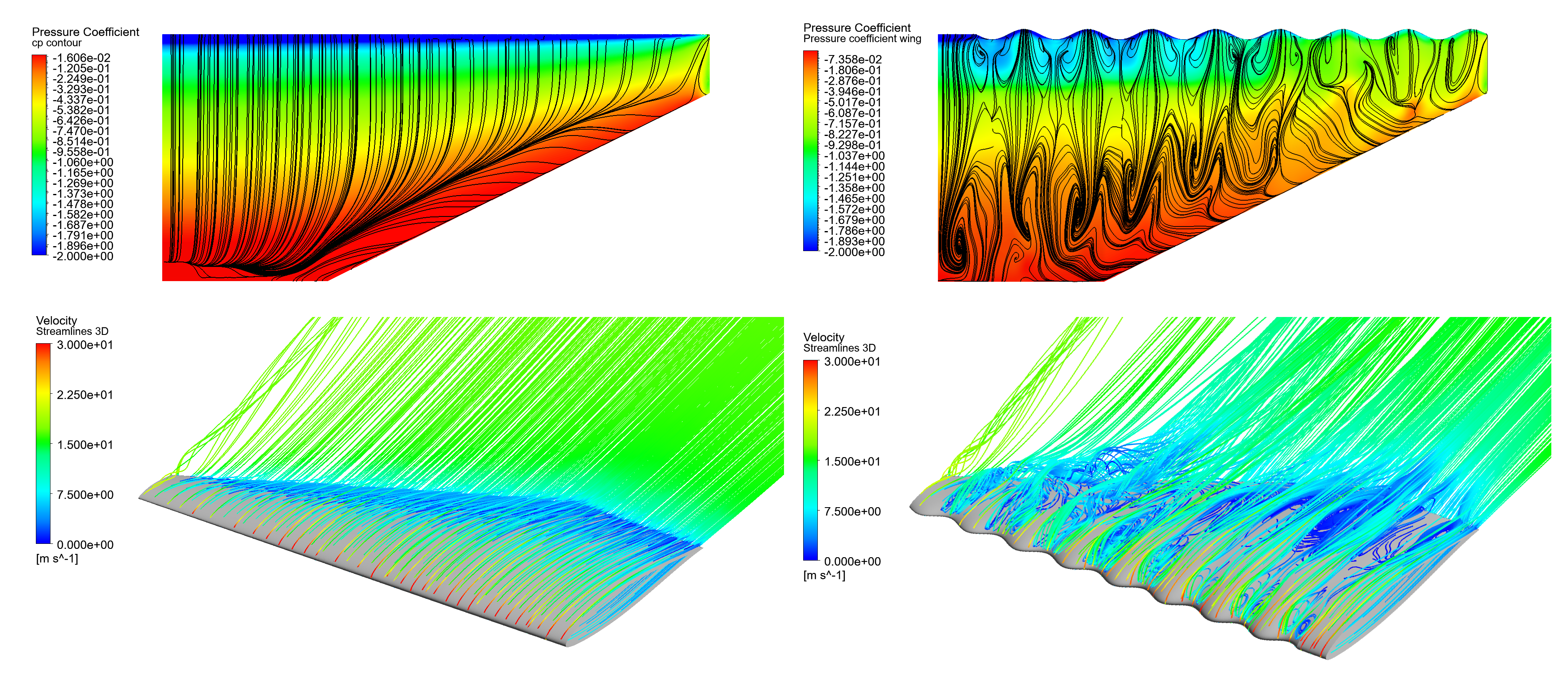}}
    \caption{Surface pressure coefficient contour and shear stress lines over the upper surface of the straight (left top) and tubercle leading-edge wing (right top) in ground-effect at 10deg angle of attack. Numerical streamline flow visualization of the straight (left bottom) and serrated (bottom right) leading-edge wing in ground-effect.}
    \label{surface_shearlines_W2_10deg}
\end{figure}

The final step in the analysis of these preliminary results is determining the effects of trailing-edge sweep combined with those of serrations at the leading edge, for the wings in-ground-effect specifically. This is achieved by comparing surface pressure distribution and shear stress lines, and visualization of the flow using numerical streamlines with velocity magnitude as the parameter of interest. These results are shown in Figure \ref{surface_shearlines_W2_10deg} at $\alpha = 10 deg$ for both wings.

Figure \ref{surface_shearlines_W2_10deg} shows, for the wing with straight leading edge, a clear spanwise suction peak at the leading-edge that decreases spanwise toward the tip of the wing, and a cross-flow directed inboard along the trailing-edge. Surface pressure distribution and shear stress lines follow the expected development for a two-element trapezoidal wing with a transition from favorable to adverse pressure gradient chordwise. These results are supported by the numerical flow visualization.

The flow over the upper surface of the wing with serrated leading-edge is visually more chaotic, resulting in a surface pressure distribution that departs from classical aerodynamics. At the inboard section, closest to the root section of the wing, and moving outboard towards the tip, results show that the presence of the pressure peak is significantly impacted, and its spanwise presence limited by the presence of the serration, with a strong pressure peak only at the peak of the first three serrations (from the root) and decreasing spanwise. In addition, a plateau of high low-pressure coefficients is present behind the suction peak at the first two serrations, unlike in the case of the wing with the straight leading edge. This behaviour is likely correlated by a combination of the generation of the pairs of counter-rotating vortices at the leading-edge and of the straight trailing-edge as this geometric feature stabilizes spanwise the pressure distribution. Interestingly, this phenomenon is also present at the first two serrations on the outboard section of the wing, where the trailing-edge sweep is already present, but where the chordwise low pressure deceleration region, in the range between $C_p \approx 4e^-1$ and $C_p \approx 1.6e^-2$, therefore extending the region of higher adverse pressure gradient.     

In the region where the effects of trailing-edge sweep induced cross-flow is more dominant, the magnitude of the pressure peak at the leading-edge is much lower and the pressure gradient variation is not as marked. Since the distance traveled by the counter-rotating vortex pairs from the serrated leading edge is shorter, compared to those on the inboard section of the wing, more energy is conserved in the flow which limits the adverse pressure gradient and affects the cross-flow pattern at the trailing-edge. 

These results provide an initial insight in the drop in coefficient of lift and increase in drag coefficient at angles of attack between $6 deg$ and $10 deg$ shown in Figure \ref{aero_coefficients_graphs}(a).
     
\section{Conclusions}\label{Sec5}
A study on the effects of a serrated leading-edge has been conducted numerically and experimentally for a two-element trapezoidal wing with a swept trailing-edge section. Numerical simulations were conducted to comparatively assess the performance of both wing geometries in two different situations: one out-of-ground-effect and another one in-ground-effect at a distance of $0.4c_{MAC}$ from the ground. Wind-tunnel tests of the two wings and in two ground clearance cases were performed at the Mark V. Morkovin wind-tunnel at Illinois Institute of Technology (IIT) to validate the numerical results. 

Results indicate, in agreement with available literature, and suggesting in maximum lift coefficient and stall angle of the wing with the serrated leading-edge compared to the wing with the straight leading-edge. An overall increase in aerodynamic performance of both wing with decrease in distance from the ground was also documented. For the case of the wings out-of-ground-effect, a clear improvement in maximum lift coefficient and stall angle has been shown for the wing with the serrated leading-edge geometry.

Spanwise contours of the pressure coefficient for the wings in ground effect at a distance of $0.4c_{MAC}$ reveal a significant variation in pressure distribution on both the suction and pressure side which enhance the overall aerodynamic efficiency at low angle. However, a detrimental effect at greater angles of attack is found as the area of strong adverse pressure gradient towards the trailing-edge is larger compared to that of a wing with a straight leading-edge, at equivalent height from the ground. The analysis of these contours at several chord distances from the leading edge also demonstrate a larger impact of the counter-rotating vortex pairs generated by the serrations at the leading-edge at the region of the wing tip.    

\section*{Acknowledgments}\label{Sec5}
The authors would like to acknowledge Mr. Craig Johnson and  Mr. Joseph Borrelli of IIT for their support in setting up the wind-tunnel test, and Prof. Dave Williams for lending us the DAQ System. The authors would also like to acknowledge the Lundeqvists Stiftelse for the financial support. R. Vineusa was supported by was supported by ERC Grant No. 2021-CoG-101043998, DEEPCONTROL. The views and opinions expressed are however those of the author(s) only and do not necessarily reflect those of European Union or European Research Council.




\bibliographystyle{elsarticle-num}
\bibliography{IIT_bibliography}







\end{document}